\documentclass[aps,prb,twocolumn,footinbib,longbibliography,superscriptaddress]{revtex4-1}
\usepackage{amsmath,amssymb}
\usepackage{bm}
\usepackage{epsfig}
\usepackage{natbib} 
\usepackage{hyperref}
\usepackage{color}
\usepackage{graphicx}

 \hypersetup{pdfstartview={FitH},pdfpagemode={UseNone},
            colorlinks,linkcolor=red, citecolor=blue, urlcolor=blue,
            bookmarks=true, bookmarksopen=true, pdfnewwindow=true}

\newcommand{\rmi}{{\rm i}}

\newcommand{\e}{{\rm e}}

\newcommand{\av}[1]{\left< #1 \right>}
\newcommand{\ST}[1]{\textcolor{magenta}{#1}}

\graphicspath{{./figs/}}

\begin{document}

\title{Spin noise at electron paramagnetic resonance}

\author{\firstname{A.~V.} \surname{Poshakinskiy}}
\email{poshakinskiy@mail.ioffe.ru}
\affiliation{Ioffe  Institute, St.~Petersburg 194021, Russia}
\author{\firstname{S.~A.} \surname{Tarasenko}}
\affiliation{Ioffe  Institute, St.~Petersburg 194021, Russia}

\begin{abstract}

We develop a microscopic theory of spin noise in solid-state systems at electron paramagnetic resonance, when the spin dynamics is driven by static and radio-frequency (RF) magnetic fields and the stochastic effective magnetic field stemming from the interaction with environment. The RF field splits the peaks in the power spectrum of spin noise into the 
Mollow-like triplets and also gives rise to additional spin-spin correlations which oscillate in the absolute time at the 
RF frequency and the double frequeqncy. Even in systems with strong inhomogeneous broadening, the spin noise spectrum contains narrow lines insensitive to the dispersion of the effective $g$-factors. Thus, the measurements of spin noise at electron paramagnetic resonance provides an access to the intrinsic spin lifetime of electrons. 

\end{abstract}
 \maketitle
 
\section{Introduction}

The optical spectroscopy of spin fluctuations~\cite{Ivchenko73fl}, first demonstrated for a vapor of 
alkali atoms~\cite{Aleksandrov1981,Crooker2004} and lately applied to solid-state systems~\cite{Oestreich2005}, has developed into a powerful method for the study of spin dynamics~\cite{Hubner2014,Zapasskii2013review} and diffusion~\cite{Poshakinskiy2016cor,Cronenberger2019}. 
While most of early experimental studies benefit from the fact than the noise spectroscopy requires no excitation of an electron system and allows one to study the system in the conditions close to thermal equilibrium, the modern spin noise spectroscopy goes beyond the measurement of linear response~\cite{Smirnov2017,Wiegand2018}. 
In particular, the study of spin noise in the presence of a radio-frequency (RF) field provides an additional insight into the physics of fluctuations and spin dephasing processes~\cite{Braun2007,Yue2015,Glasenapp2014,Brox2011,Sinitsyn2016,Tarasenko2018}. 
A scheme to measure the spin fluctuations via optical Faraday rotation with an oscillating magnetic field instead of a static magnetic field commonly used in experiments was discussed in Ref.~\onlinecite{Braun2007}. The power spectrum of electron spin noise driven by a RF field for an ensemble of localized electrons in the presence of a random hyperfine field was theoretically studied in Ref.~\onlinecite{Yue2015}. Experimentally, the spin noise spectroscopy in the geometry of electron paramagnetic resonance with a static magnetic field and a weak perpendicular RF field was recently realized for a vapor of potassium atoms demonstrating an access to the spin dynamics beyond thermal equilibrium and linear response~\cite{Glasenapp2014}.  
An approach to probe spin quadrupole noise of spin-3/2 color centers in semiconductors by measuring optical blinking was proposed in Ref.~\onlinecite{Tarasenko2018}.
 
Here, we present a microscopic theory of electron spin fluctuations in the conditions of paramagnetic resonance
when electrons interacting with environment are subject to a static magnetic field and a weak perpendicular RF magnetic field. We obtain analytical expressions for the components of spin-spin correlation function and study the effect of  
inhomogenious broadening on the noise spectrum. We show how the RF field suppresses the spin dephasing caused by the dispersion of the effective $g$-factors thereby providing an access to the intrinsic spin lifetime of electrons. 
We also demonstrate that the RF field driven spin correlations contain contributions oscillating with the 
absolute time at the single and double frequencies of the RF field. These contributions are accessible by a lock-in technique.

\section{Theory}

We consider the spin dynamics of an ensemble of electrons in the static magnetic field $\bm B \parallel z$ and a weak perpendicular circularly polarized RF magnetic field $\bm B_{\rm rf}(t)$, see Fig.~\ref{fig:model}. Interaction with environment for a spin-1/2 particle is described in terms of a fluctuating effective magnetic field $\bm B_{\rm fl}(t)$ acting on the particle spin. Thus, the dynamics of the spin operator $\bm{s}$ of an individual electron in the Heisenberg picture is governed by the equation
\begin{equation}\label{dsdt}
\frac{d \bm s}{d t} = [\bm\Omega_L +\bm\Omega_{\rm rf}(t) + \bm\Omega_{\rm fl}(t)] \times \bm s \,,
\end{equation}
where $\bm\Omega_L = (0,0, \Omega_L)$, $\bm\Omega_{\rm rf}(t) = \Omega_{\rm rf} (\cos \omega_{\rm rf} t, \sin \omega_{\rm rf} t,0)$, and 
$\bm\Omega_{\rm fl}(t)$ are the Larmor frequencies corresponding to the fields $\bm B$, $\bm B_{\rm rf}(t)$, and $\bm B_{\rm fl}(t)$, respectively.

\ST{
\begin{figure}[t]
\includegraphics[width=0.5\columnwidth]{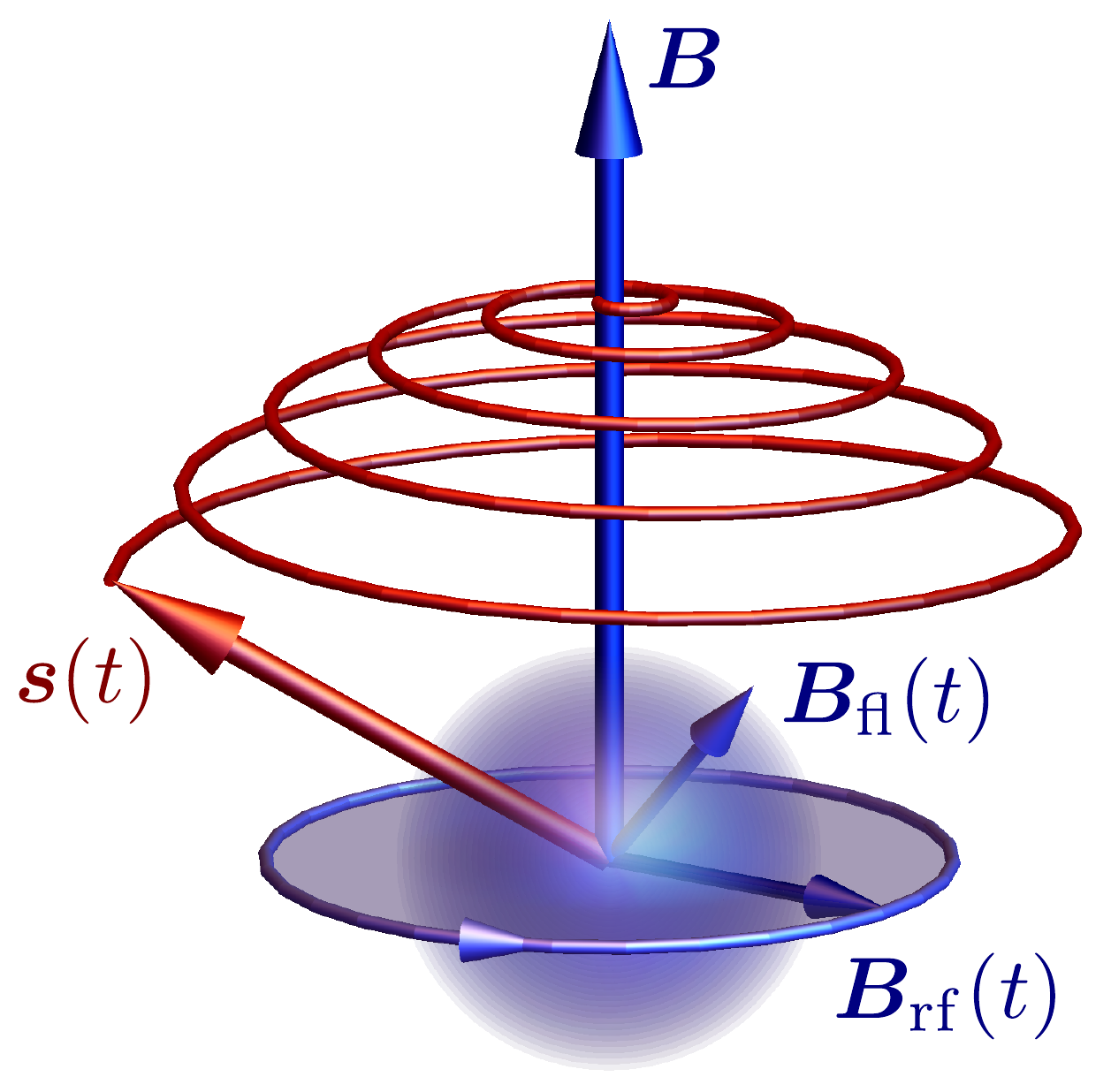}
\caption{Dynamics of an electron spin $\bm s$ in the presence of static $\bm B$, radio-frequeqncy $\bm B_{\rm rf}(t)$, and fluctuating $\bm B_{\rm fl}(t)$ magnetic fields.} 
\label{fig:model}
\end{figure}
}

The spin correlation function, which is a measure of spin fluctuations, is defined by
\begin{equation}\label{K_gen}
K_{\alpha\beta}(t,t') = \av{ \{ s_\alpha(t) , s_\beta(t') \} } ,
\end{equation}
where $\{ s_\alpha(t) , s_\beta(t') \} = [s_\alpha(t) s_\beta(t') + s_\beta(t') s_\alpha(t)]/2$ is the symmetrized product and the angle brackets denote averaging with the spin-density matrix of electrons. Below, we calculate the correlation function for different regimes of spin dynamics.

\subsection{Spin noise in the absence of static and RF fields}

First, we describe briefly the procedure to calculate spin noise in the absence of static and RF fields. In this case, the spin dynamics of an electron is described by the equation
\begin{equation}\label{dsdt_simp}
\frac{d \bm s}{d t} = \bm\Omega_{\rm fl}(t) \times \bm s \,.
\end{equation}

The formal solution of Eq.~\eqref{dsdt_simp} can be presented in the form
\begin{equation}
\bm s(t) = \bm U(t,0) \bm s(0) \,,
\end{equation}
where $\bm U(t,0)$ is the matrix describing the spin evolution. The expansion of $\bm U(t,0)$ into the series in $\bm \Omega_{\rm fl} t$
yields 
\begin{align}\label{U}
\bm U(t,0) \bm s(0) &= \Big[ 1 + \int\limits_{0}^{t} \bm \Omega_{\rm fl} (t_1) dt_1 \times \\
&+ \int\limits_{0}^{t} \bm \Omega_{\rm fl} (t_1) dt_1 \times \int\limits_{0}^{t_1} \bm \Omega_{\rm fl} (t_2) dt_2 \times  + \ldots \Big] \bm s(0) \nonumber \,.
\end{align}
The spin correlation function~\eqref{K_gen} is then given by
\begin{equation}
K_{\alpha \beta}(t,t') = \sum_{\gamma, \delta} \av{ \{ U_{\alpha \gamma}(t,0) s_{\gamma}(0) ,  U_{\beta \delta}(t',0) s_{\delta}(0) \} } .
\end{equation}
Taking into account that, for spin-1/2 particles, $\{ s_\alpha(t) , s_\beta(t)\} = \delta_{\alpha \beta} /4$ at any $t$ we obtain
\begin{equation}\label{K_gen2}
\bm K(t,t') = \frac{ \av{ \bm U(t,t') } \theta(t-t') + \av{ \bm U^T(t',t) } \theta(t-t') }{4} \,.
\end{equation}

We assume that the mean value of the effective magnetic field $\bm B_{\rm fl}(t)$ originating, e.g., from electron-phonon interaction or electron motion in a non-centrosymmetric structure,
is zero. Besides, the effective field is weak
and rapidly fluctuates in time so that the condition $\Omega_{\rm fl} \tau_c \ll 1$ is satisfied, where $\tau_c$ is the correlation time
of the effective field. In this case, the spin relaxation time of the electron ensemble $T_s$ is of the order of $1/(\Omega_{\rm fl}^2 \tau_c)$ being much longer than $\tau_c$~\cite{Dyakonov08:SPS}. On the timescale $\delta t$ much smaller than the spin relaxation time $T_s$ but much
larger than the correlation time $\tau_c$, the ensemble-averaged matrix of the spin evolution reads
\begin{equation}\label{dU}
\av{ U_{\alpha \beta}(t + \delta t , t) } = \delta_{\alpha \beta} - \Gamma_{\alpha \beta}(t) \delta t \,,
\end{equation}
where $\Gamma_{\alpha\beta}(t)$ is the tensor 
defined by
\begin{equation}\label{Gamma_gen}
\Gamma_{\alpha\beta}(t) = \int\limits_0^\infty \langle \bm\Omega_{\rm fl}(t-\tau) \cdot \bm\Omega_{\rm fl}(t) \delta_{\alpha\beta} 
- \Omega_{{\rm fl},\alpha}(t-\tau) \Omega_{{\rm fl},\beta}(t) \rangle d\tau .
\end{equation}
Equations~\eqref{dU} and~\eqref{Gamma_gen} which follow from Eq.~\eqref{U}.

Equations~\eqref{K_gen2} and~\eqref{dU} allow us to formulate the differential equations for the spin correlation function
\begin{align}\label{K_diff}
\frac{\partial \bm K(t,t')}{\partial t} &= \bm \Gamma^T(t) \bm K(t,t') \theta(t'-t)
 - \bm \Gamma(t) \bm K(t,t') \theta(t-t') , \nonumber \\
\frac{\partial \bm K(t,t')}{\partial t'} &\hspace{-.05cm}=\hspace{-.05cm} \bm K(t,t') \bm\Gamma(t') \theta(t-t')
 \hspace{-.05cm}-\hspace{-.05cm} \bm K(t,t') \bm\Gamma^T(t') \theta(t'-t) ,
\end{align}
which must be solved with the initial condition $K_{\alpha \beta}(t,t) = \delta_{\alpha \beta}/4$. Equations~\eqref{K_diff} are very general equations for the correlation function of an ensemble of spin-1/2 particles. Below, we use they to calculate the spectra of spin noise in various conditions. 

In an isotropic ergodic system in the absence of external fields, the tensor $\Gamma_{\alpha \beta}$ is diagonal, $\Gamma_{\alpha \beta} = \gamma_0 \delta_{\alpha \beta}$, and independent of time. It has the meaning of the tensor of spin relaxation rates. 
Solution of Eqs.~\eqref{K_diff} for this simple case has the form
\begin{equation}\label{K_final1}
K_{\alpha \beta} (t,t') = \frac{\delta_{\alpha\beta}}{4} \exp (-\gamma_0 |t-t'|)  \,.
\end{equation}

The spin correlation function~\eqref{K_final1} is well-known and can be obtained by different theoretical approaches including the use of the fluctuation-dissipation theorem and the method of Langevin forces~\cite{Ivchenko73fl,Sinitsyn2016}. An advantage of the method presented above is that it enables the straightforward calculation of the correlation function in the presence of external driving forces.

\subsection{Spin noise in a static magnetic field}\label{sec:static}

In the presence of a static magnetic field $\bm B \parallel z$, the spin dynamics of an electron
can be fruitfully analyzed in the coordinate frame $(x',y',z')$ rotating about the $z$ axis with the angular frequency $\Omega_L$. Mathematically, this coordinate frame change is equivalent to the substitution $\bm s = \bm R \, {\bm s}'$ and $\bm \Omega_{\rm fl} = \bm R \, {\bm \Omega}'_{\rm fl}$ in Eq.~\eqref{dsdt}, where $\bm R$ is the matrix of rotation
\begin{equation}\label{Rotation_matrix}
\bm R(t) = \left( 
\begin{array}{ccc} 
\cos\Omega_L t  & -\sin \Omega_L t & 0 \\ 
\sin \Omega_L t & \cos \Omega_L t & 0 \\  
0 & 0 & 1
\end{array} 
\right) .
\end{equation}
In the rotating coordinate frame, the equation of spin dynamics has the form
\begin{equation}\label{dsdt_rot}
\frac{d \bm s'}{d t} = \bm\Omega_{\rm fl}'(t) \times \bm s' 
\end{equation}
which is equivalent to Eq.~\eqref{dsdt_simp}. Therefore, the spin correlation function in the rotating frame $\bm K'(t,t')$
satisfies Eqs.~\eqref{K_diff} with the tensor $\bm \Gamma'$ determined by Eq.~\eqref{Gamma_gen} where $\bm \Omega_{\rm fl}(t)$ is replaced by $\bm \Omega'_{\rm fl}(t) = \bm R^{-1}(t) \bm \Omega_{\rm fl}(t)$.

To be specific we assume that the fluctuating field $\bm \Omega_{\rm fl}(t)$ is described by the correlation function
\begin{align}\label{OfOf}
\av{\Omega_{{\rm fl},\alpha}(t) \, \Omega_{{\rm fl},\beta}(t')} = \frac{\Omega_{\rm fl}^2}3  \exp(-|t-t'|/\tau_c)  \delta_{\alpha\beta} \,.
\end{align}
where $\Omega_{\rm fl}$ is the characteristic amplitude of the field and $\tau_c$ is the correlation time. Then, the tensor $\bm \Gamma'$ has the form
\begin{align}\label{Gamma_rot}
\bm \Gamma' = \frac{\Omega_f^2}3 \int\limits_0^\infty \left[ \bm I \, {\rm Tr\,} \bm R(\tau)  - \bm R(\tau)  \right]  \e^{-\tau/\tau_c} d\tau ,
\end{align}
where $\bm I$ is the $3\times 3$ identity matrix and we took into account that the matrix of rotation is orthogonal, 
i.e., $\bm R^{-1} = \bm R^{T}$. 

Straightforward calculation of the integral in Eq.~\eqref{Gamma_rot} with the rotation matrix~\eqref{Rotation_matrix} shows that the tensor $\bm \Gamma'$ has both the symmetric $\bm \Gamma'^{\rm (s)} = (\bm \Gamma' + \bm \Gamma'^T)/2$ and the antisymmetric $\bm \Gamma'^{\rm (a)} = (\bm \Gamma' - \bm \Gamma'^T)/2$ parts. The symmetric part stands for the tensor of spin relaxation rates. Its non-zero components are 
\begin{equation}
\Gamma'^{\rm (s)}_{x'x'} = \Gamma'^{\rm (s)}_{y'y'} = \gamma_{\perp}\,, \;\; \Gamma'^{\rm (s)}_{z'z'} = \gamma_z \,,
\end{equation}
where
\begin{align}\label{gamma_rates}
\gamma_{\perp} = \frac{\gamma_0}2\left( 1+\frac{1}{1+\Omega_L^2\tau_c^2}\right) \,, \ 
\gamma_z = \frac{\gamma_0}{1+\Omega_L^2\tau_c^2} \,,
\end{align}
and $\gamma_0=(2/3)\Omega_{\rm fl}^2 \tau_c$. The antisymmetric second-rank tensor $\bm \Gamma'^{(a)}$ is equivalent to a pseudovector
$\delta \bm \Omega_L$ and represents, in fact, a stochastic-field-induced correction to the Larmor frequency $\bm \Omega_L$. For the isotropic case under study, $\delta \bm \Omega_L \parallel \bm \Omega_L$ and has the form
\begin{equation}\label{Omega_corr}
\delta \bm \Omega_L = \frac{\gamma_0}2 \, \frac{\bm \Omega_L\tau_c}{1+\Omega_L^2\tau_c^2} .
\end{equation}
The correction to the Larmor frequency can also be interpreted as a (magnetic field dependent) renormalization of the effective electron $g$-factor. The renormalization of the effective $g$-factor caused by the spin-orbit effective magnetic field at the cyclotron motion of electrons in three-dimensional and two-dimensional semiconductor structures was analyzed in Refs.~\cite{Ivchenko73,Tarasenko09,Poshakinskiy11a}. In our case, the renormalization occurs due to the action of the stochastic effective magnetic field on the electron spin.

Solution of Eqs.~\eqref{K_diff} for the tensor $\bm \Gamma'$ independent of time gives 
the spin correlation function in the rotating coordinate frame
\begin{align}\label{Krot}
\bm K'(t,t') = \frac{1}{4} \exp \left[ -\bm\Gamma'^{\rm (s)} |t-t'| -\bm\Gamma'^{\rm (a)}(t-t') \right] .
\end{align}

The spin correlation function in the initial (fixed) coordinate frame $\bm K(t,t')$ is readily obtained 
by the transformation $\bm K(t,t') = \bm R(t) \bm K'(t,t') \bm R^{-1}(t')$.  The corresponding calculations give
\begin{align}\label{KB}
&K_{xx}(t,t')= K_{yy}(t,t') = \frac{1}{4} \cos [\tilde\Omega_L(t-t')]\,\e^{-\gamma_\perp|t-t'|} \,,\nonumber\\
&K_{xy}(t,t')= - K_{yx}(t,t') = - \frac{1}{4} \sin [\tilde\Omega_L(t-t')]\,\e^{-\gamma_\perp|t-t'|} \,,\nonumber\\
&K_{zz}(t,t')= \frac{1}{4}\,\e^{-\gamma_z |t-t'|} \,,
\end{align}
where $\tilde\Omega_L = \Omega_L+\delta\Omega_L$ is the Larmor frequency renormalized by the stochastic field.
The spin relaxation rates $\gamma_\perp$ and $\gamma_z$ and the frequency correction $\delta \Omega_L$ are given by Eqs.~\eqref{gamma_rates} and~\eqref{Omega_corr}, respectively.

The spectrum of spin noise, which is usually investigated in experiments, is determined by the Fourier image of the spin correlation function $\bm K(t,t') = \int \bm K(\omega)\e^{-\rmi\omega (t-t')} d \omega/2 \pi$. The corresponding Fourier images of the functions~\eqref{KB} have the form 
\begin{align}\label{KB_Fourier}
&K_{xx}(\omega)= K_{yy}(\omega) = \frac14 \sum_{\pm} \frac{\gamma_{\perp}}{(\omega \pm \tilde\Omega_L)^2 + \gamma_{\perp}^2}  \,,\nonumber\\
&K_{xy}(\omega)= - K_{yx}(\omega) = \frac{\rmi}{4} \sum_{\pm} \frac{\pm \gamma_{\perp}}{(\omega \pm \tilde\Omega_L)^2 + \gamma_{\perp}^2} \,,\nonumber\\
&K_{zz}(\omega)= \frac12 \frac{\gamma_z}{\omega^2 + \gamma_z^2} \,.
\end{align}

Equations~\eqref{KB} and~\eqref{KB_Fourier} generalize previous results for the spin dynamics and spin noise of electrons in an external static magnetic field~\cite{Aleksandrov1981,Braun2007}. They show that, even in isotropic systems, stochastic effective magnetic field can modify the electron $g$-factor and leads to an anisotropy of spin relaxation.

\begin{figure*}[t]
\includegraphics[width=0.85\textwidth]{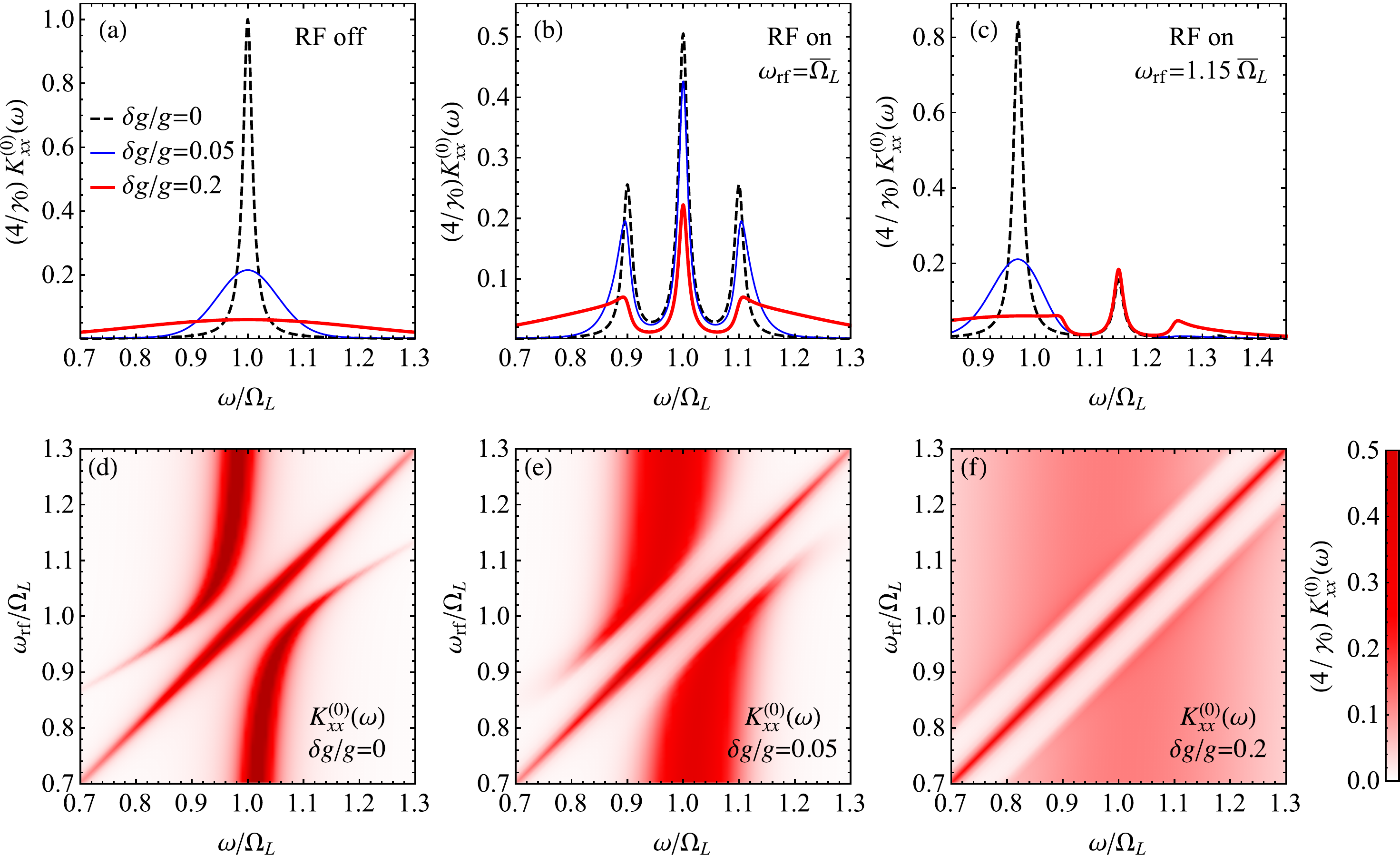}
\caption{
Noise spectra of the perpendicular spin component $K_{xx}^{(0)}(\omega)$. Panel (a) shows the spectra in the absence of  RF field for various value of $g$-factor dispersion $\delta g$.  Panels (b) and (c) correspond to the case when the RF field with the frequency $\omega_\text{rf}$ indicated in the graphs and the intensity corresponding to the Rabi frequency $\Omega_\text{rf} = 0.1\Omega_L$ is turned on. Panels (d)-(f) show the evolution of the spectra when the frequency $\omega_\text{rf}$ is scanned. The spin relaxation rate was chosen to be $\gamma_0 = 0.01\Omega_L$. 
} 
\label{fig:K0xx}
\end{figure*}

\begin{figure*}[t]
\includegraphics[width=0.85\textwidth]{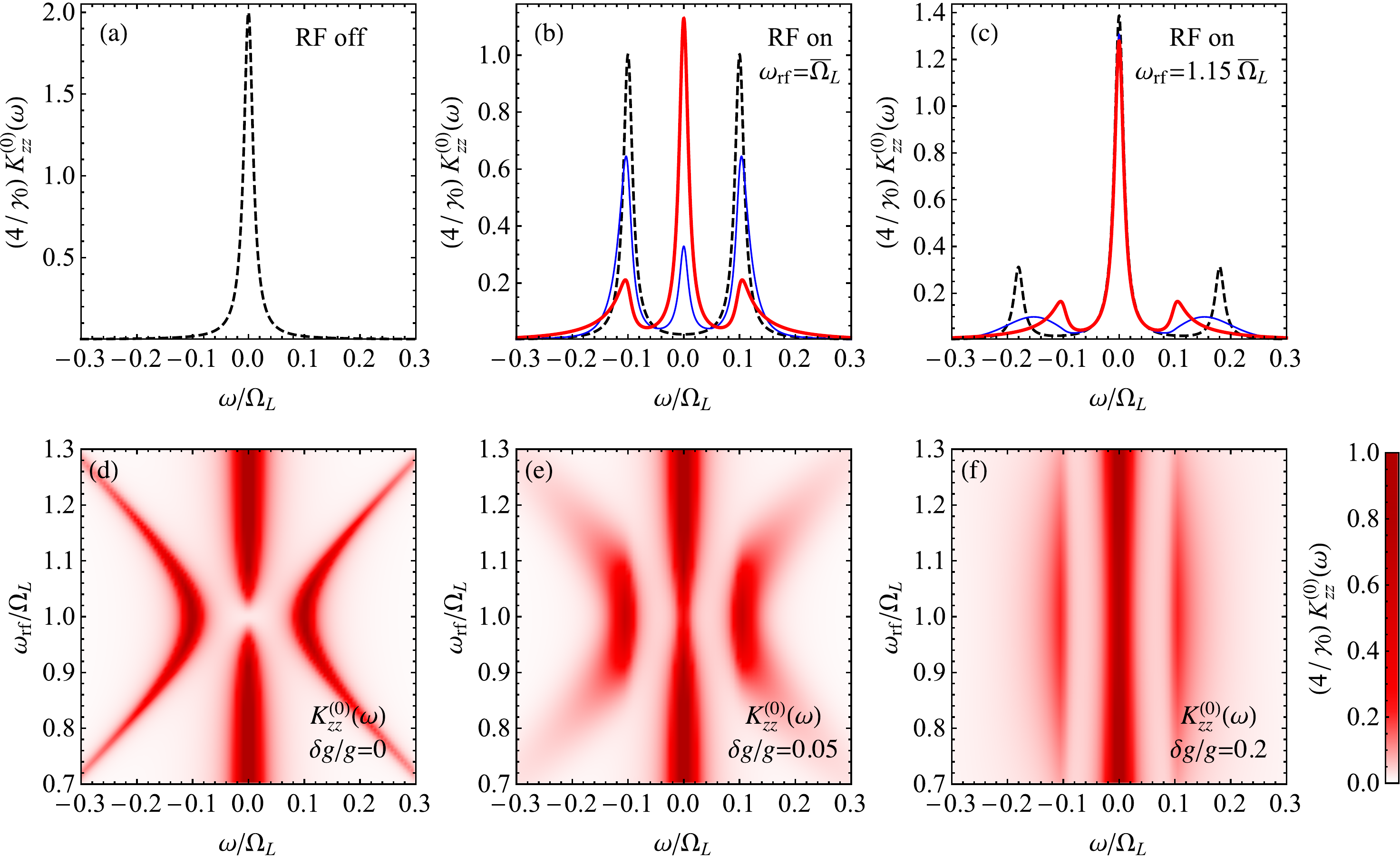}
\caption{
Noise spectra of the longitudinal spin component $K_{zz}^{(0)}(\omega)$. Panel (a) shows the spectra in the absence of  RF field for various value of $g$-factor dispersion $\delta g$.  Panels (b) and (c) correspond to the case when the RF field with the frequency $\omega_\text{rf}$ indicated in the graphs and the intensity corresponding to the Rabi frequency $\Omega_\text{rf} = 0.1\Omega_L$ is turned on. Panels (d)-(f) show the evolution of the spectra when the frequency $\omega_\text{rf}$ is scanned.  The spin relaxation rate was chosen to be $\gamma_0 = 0.01\Omega_L$. 
} 
\label{fig:K0zz}
\end{figure*}

\subsection{Spin noise in static and RF fields}

Now we are ready to discuss the spin noise in the presence of both the static magnetic field $\bm B \parallel z$ and 
the RF field $\bm B_{\rm rf}(t) \perp z$. The RF field rotates about the $z$ axis with the frequency $\omega_{\rm rf}$.
Therefore, it is convenient to consider the spin dynamics in the coordinate frame $F'$ rotating with the RF field. In this coordinate frame, the equation of spin dynamics has the form
\begin{align}\label{dsdt_rot1}
\frac{d\bm s'}{dt} = [\bm\Omega' + \bm\Omega_{\rm fl}'(t)] \times \bm s' \,,
\end{align}
where $\bm\Omega' = (\Omega_{\rm rf},0,\Omega_L-\omega_{\rm rf})$, $\bm s' = \bm R_1^{-1} \bm s$, 
$\bm\Omega_{\rm fl}' =  \bm R_1^{-1} \bm\Omega_{\rm fl}$, and $\bm R_1$ is the matrix of rotation. 
The matrix $\bm R_1(t)$ is given by Eq.~\eqref{Rotation_matrix} where $\Omega_L$ is replaced by $\omega_{\rm rf}$.

Equation~\eqref{dsdt_rot1} describes the spin dynamics in the static and stochastic magnetic fields. To solve this equation and calculate the spin correlation function we apply the method developed in Sec.~\ref{sec:static}. First, we change once again the coordinate frame
and consider the spin dynamics in the frame $F''$ rotating with the frequency $\bm\Omega'$ with respect to the frame $F'$.  This transformation is described by the rotation matrix 
\begin{equation}\label{Rotation_matrix2}
\bm R_2(t) = \left( 
\begin{array}{ccc} 
u^2 + v^2 \cos\Omega' t  & -v \sin \Omega' t & uv(1-\cos \Omega' t) \\ 
v \sin \Omega' t & \cos \Omega' t & - u \sin \Omega't \\  
uv (1-\cos \Omega't) & u \sin \Omega't & v^2 + u^2 \cos \Omega't
\end{array} 
\right) ,
\end{equation}
where $\Omega'=|\bm \Omega'| = \sqrt{\Omega_{\rm rf}^2 + (\Omega_L - \omega_{\rm rf})^2}$, $u = \Omega_{\rm rf}/\Omega'$, 
and $v = (\Omega_L - \omega_{\rm rf}) / \Omega'$.  In the frame $F''$, the evolution of the spin operator $\bm s'' =  \bm R_2^{-1}(t) \bm R_1^{-1}(t) \bm s$ is governed solely by the fluctuating field $\bm\Omega_{\rm fl}''(t) = \bm R_2^{-1}(t) \bm R_1^{-1}(t) \bm\Omega_{\rm fl}(t)$. Then, we calculate the tensor $\bm\Gamma''$ in the frame $F''$ using Eq.~\eqref{Gamma_gen} where $\bm\Omega_{\rm rf}(t)$ is replaced by $\bm\Omega_{\rm rf}''(t)$. Because the rotation matrices $\bm R_1$ and $\bm R_2$ do not commute with each other, 
the tensor $\bm \Gamma''$ in general case becomes time-dependent. This time dependence is given by $\bm\Gamma''(t) = \bm R_2^T(t) \bm \Gamma' \bm R_2(t)$ with time-independent $\bm\Gamma'$. For the correlation function~\eqref{OfOf}, the tensor $\bm \Gamma'$ has the form
\begin{align}\label{Gamma1}
\bm\Gamma' = \frac{\Omega_{\rm fl}^2}3 \int\limits_0^\infty \left\{ \bm I\, {\rm Tr} [\bm R_2(\tau)\bm R_1(\tau)]  - \bm R_2(\tau)\bm R_1(\tau)  \right\}  \e^{-\tau/\tau_c} d\tau .
\end{align}

To simplify further calculations we consider that the correlation time of the fluctuating field $\tau_c$ is short in comparison with 
$1/\Omega'$ and $1/\omega_{\rm rf}$. In this regime, one has $\bm \Gamma'' = \bm \Gamma' = \bm I \gamma_0$ and, accordingly,
\begin{equation}
\bm K''(t,t') =  \frac{\exp (-\gamma_0 |t-t'|)}{4}  \bm I \,.
\end{equation}

Finally, the spin correlation function in the fixed coordinate frame has the form
\begin{align}\label{K_F0}
\bm K(t,t') = \bm R_1(t) \bm R_2(t) \bm K''(t,t') \bm R_2^T(t') \bm R_1^T(t') \,.
\end{align}

The presence of an external RF field breaks the time homogeneity in the system. Therefore, the correlation function
$\bm K(t,t')$ depends not only on the time difference $t-t'$ but also on $(t+t')/2$. The latter dependence is periodic
in $2\pi/\omega_{\rm rf}$ because of the RF field periodicity. 

To analyze the spectrum of spin noise we expand the correlation function in the Fourier series as follows
\begin{align}\label{Ktw}
\bm K(t,t') = \sum_{n} \int \frac{d\omega}{2\pi} \bm K^{(n)}(\omega) \e^{-\rmi n \omega_{\rm rf}(t+t')/2 -\rmi\omega (t-t')} \,.
\end{align}
The function $ \bm R_2(t) \bm K''(t,t') \bm R_2^T(t')$ depends only on the time difference $t-t'$ and the 
rotation matrices $\bm R_1(t)$ and $\bm R_1^T(t)$ contain only the first time harmonics. Therefore, the only nonzero harmonics $\bm K^{(n)}(\omega)$ are those with $n=0,\pm1,\pm2$, and $\bm K^{(-n)}(\omega) = \bm K^{(n)*}(-\omega)$.

The harmonics $\bm K^{(0)}(\omega)$, which are usually studied in experiments, have the form
\begin{align}\label{eq:Kw}
&K^{(0)}_{zz} (\omega) =  \frac{(\Omega_L-\omega_{\rm rf})^2}{\Omega'^2} \Delta_+(\omega,0)+ \frac{\Omega_{\rm rf}^2}{\Omega'^2}  \Delta_+(\omega,\Omega') , \\
&K^{(0)}_{xx} (\omega) = K^{(0)}_{yy}(\omega) = \frac{(\Omega_L-\omega_{\rm rf} + \Omega')^2}{4\Omega'^2}  \Delta_+ (\omega,\omega_{\rm rf}+\Omega')  \nonumber \\ \nonumber
&+ \frac{(\Omega_L-\omega_{\rm rf} - \Omega')^2 }{4\Omega'^2}\Delta_+ (\omega,\omega_{\rm rf}-\Omega') 
+ \frac{\Omega_{\rm rf}^2}{2\Omega'^2} \Delta_+(\omega, \omega_{\rm rf} ) ,  \\
&\rmi K^{(0)}_{yx} (\omega) = -\rmi K^{(0)}_{xy} = \frac{(\Omega_L-\omega_{\rm rf} + \Omega')^2 \Delta_- (\omega,\omega_{\rm rf}+\Omega')}{4\Omega'^2} \nonumber\\\nonumber
& + \frac{(\Omega_L-\omega_{\rm rf} - \Omega')^2 \Delta_- (\omega,\omega_{\rm rf}-\Omega')}{4\Omega'^2} + \frac{\Omega_{\rm rf}^2 \Delta_- (\omega,\omega_{\rm rf})}{2\Omega'^2} , 
\end{align}
where
\begin{align}\label{eq:Oprime}
\Omega' = \sqrt{\Omega_{\rm rf}^2+(\Omega_L-\omega_{\rm rf})^2}
\end{align}
 and
\[
\Delta_\pm(\omega,\omega_0) = \frac{1}{4} \left[ \frac{\gamma_0}{(\omega+\omega_0)^2+\gamma_0^2} \pm 
\frac{\gamma_0}{(\omega-\omega_0)^2+\gamma_0^2} \right] .
\]
The second term in the expression for $K_{zz}^{(0)}(\omega)$ describes the correlations of spin fluctuations centered 
at the frequency $\Omega'$ which are caused by the Rabi oscillations.
 
Correlations of the longitudinal and transverse spin components do not show up in the functions $\bm K^{(0)}(\omega)$.
However, they are revealed in the first time harmonics of the correlation function which are given by 
\begin{align}
&K^{(1)}_{xz} (\omega) = -\rmi K^{(1)}_{yz} (\omega)  =\frac{\Omega_{\rm rf}}{2\Omega'} \left\{ \Delta_-\left(\omega-\frac{\omega_{\rm rf}}2,\Omega' \right) \right. \\ \nonumber
&\left.+\frac{\Omega_L-\omega_{\rm rf}}{\Omega'} \left[\Delta_+\left(\omega-\frac{\omega_{\rm rf}}2,0\right) + \Delta_+\left(\omega-\frac{\omega_{\rm rf}}2,\Omega'\right)\right]\right\} 
\end{align}
and $K_{\alpha \beta}^{(1)}(\omega) = K_{\beta \alpha}^{(1)}(-\omega)$.

At last, the second time harmonics of the spin correlation function have the form
\begin{align}
&K^{(2)}_{xx} (\omega)= -K^{(2)}_{yy} (\omega)= -\rmi K^{(2)}_{xy} (\omega)= -\rmi K^{(2)}_{yx} (\omega) \nonumber\\
&=\frac{\Omega_{\rm rf}^2}{4\Omega'^2} \left[ \Delta_+(\omega,0) - \Delta_+(\omega,\Omega')\right] .
\end{align}

In a typical cw experiment on spin noise measurement, one collects the time-average spectrum given by $\bm K^{(0)}(\omega)$.
We note that other harmonics $\bm K^{(n)}(\omega)$ can be also extracted from the measured signals $S_{\alpha}(t)$ by performing the
averaging of the product $S_\alpha(t) S_\beta(t') \e^{\rmi n \omega_{\rm fr}(t+t')/2}$ over $t-t'$ or by applying the lock-in technique 
synchronized with the RF field. 

\section{Results and Discussion}

\begin{figure*}
\includegraphics[width=0.85\textwidth]{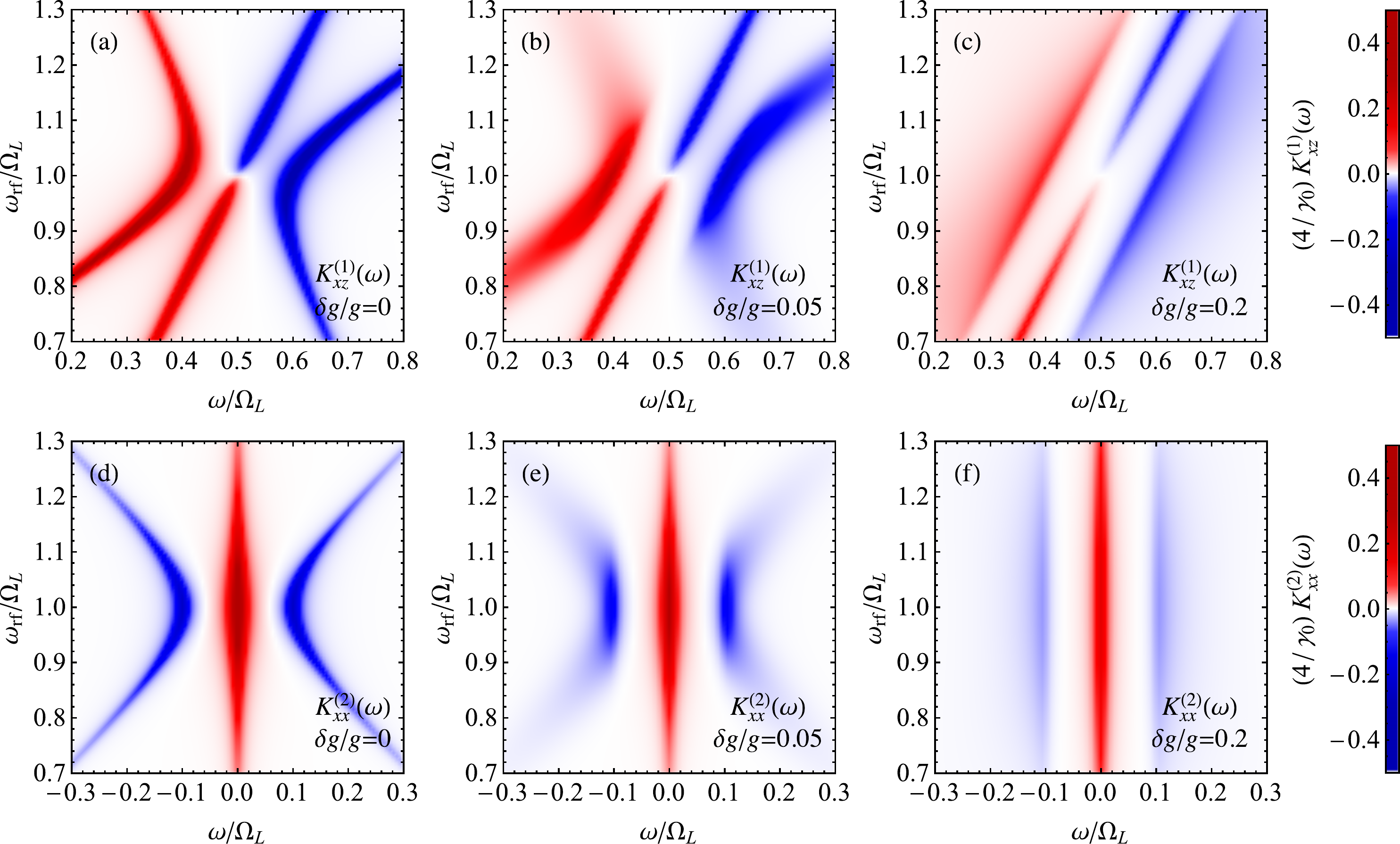}
\caption{Maps of the spin-noise correlation functions  (a)-(c) $K_{xz}^{(1)}$ and (d)-(f) $K_{xx}^{(2)}$ which arise in the presence of RF field and oscillate with absolute time at the frequencies $\omega_\text{rf}$ and $2\omega_\text{rf}$, respectively. Plots are calculated for the RF  intensity corresponding to the Rabi frequency $\Omega_\text{rf} = 0.1\Omega_L$, the spin relaxation rate $\gamma_0 = 0.01\Omega_L$, and different $g$-factor dispersion.}
\label{fig:K12}
\end{figure*}

In most of spin-noise experiments, an ensemble of spins rather than a single spin is probed. The spin ensemble
is characterized by some variation in the parameters, which leads to inhomogeneous broadening of the peaks in the spin-noise spectra. The application of RF field can suppress this broadening providing an access to the intrinsic spin relaxation time. To illustrate it, we consider the ensemble with a distribution of the effective $g$-factors resulting in a variation of the Larmor frequency $\Omega_L$. The variation of the Rabi frequency $\Omega_{\rm rf}$ is small and neglected since 
$\Omega_{\rm rf} \ll \Omega_L$. Then, for the Gaussian distribution, the spin correlation function averaged over the ensemble is given by
\begin{align}\label{K_aver}
\langle \bm K \rangle = \int \bm K \exp \left[- \frac{(\Omega_L-\bar\Omega_L)^2}{2 (\delta\Omega_L)^2} \right] \frac{d\Omega_L}{\sqrt{2\pi} (\delta\Omega_L)} \,,
\end{align}
where $\bar\Omega_L = \bar g \mu_B B_z/\hbar$, $\delta \Omega_L =  \delta g \mu_B B_z/\hbar$, $\bar g$ is the mean value of the $g$-factor, and $\delta g$ is the root-mean-square deviation.

Figure~\ref{fig:K0xx} shows the spectra of the spin-noise density $\langle K_{xx}^{(0)} \rangle$ calculated for the ensembles with different $g$-factor dispersions after Eqs.~\eqref{eq:Kw} and~\eqref{K_aver}.  Panel (a) shows the spin-noise spectra in the absence of the RF field, which feature a single peak at the average Larmor frequency of the ensemble $\bar\Omega_L$. The dispersion of the $g$-factors results in a significant broadening of the peak and a decrease of its amplitude. When the resonant RF field is on, the single peak is split into the Mollow triplet, see Fig.~\ref{fig:K0xx}(b) Importantly, the central peak of the triplet at $\omega = \omega_{\rm rf}$ is not broadened with the increase of the $g$-factor dispersion while its amplitude exhibits only a slight decrease. 
The side peaks of the triplet are shifted from the central peak by the frequency 
$\Omega' = \sqrt{\Omega_{\rm rf}^2 + (\Omega_L - \omega_{\rm rf})^2}$. In the ensemble with the dispersion of 
$\Omega_L$, the side peaks are asymmetric: their outer wings are broadened while the inner wings, which are shifted from the central peak by $\Omega_\text{rf}$, remain sharp. Therefore, the narrow peaks with the widths determined by the intrinsic spin relaxation time are resolved in the RF driven noise spectrum even for the ensembles with large dispersions while the noise spectrum in the absence of RF field is completely smeared, cf. thick red lines in Fig.~\ref{fig:K0xx}(a) and~(b). The color maps in Fig.~\ref{fig:K0xx}(d)-(f) show evolution of the spin-noise spectra when the frequency of the RF field is scanned (vertical axis). The Mollow triplet in the spectra originates from the avoided crossing of resonance line at frequency $\omega=\Omega_L$, with the lines at $\omega=\omega_{\rm rf}$ and $\omega=2\omega_{\rm rf}-\Omega_L$, see panel (d). The increase of the $g$-factor dispersion leads to smearing of lines at $\Omega_L$ and $2\omega_{\rm rf}-\Omega_L$. However, the line at $\omega_{\rm rf}$ remains narrow and the sharp fringes at $\omega_{\rm rf} \pm \Omega_{\rm rf}$ arise, see panels (e) and~(f).

The spin-noise spectra $\langle K_{zz}^{(0)} \rangle$ are shown in Fig.~\ref{fig:K0zz}. In the absence of RF field [Fig.~\ref{fig:K0zz}(a)], the noise spectrum features a peak at $\omega = 0$. The peak is not sensitive to the $g$-factor dispersion since the fluctuations of the $z$ spin component do not experience precession in the static magnetic field $\bm B \parallel z$. However, the experimental study of spin noise around zero frequency may face difficulties due to the presence of other sources of low-frequency noise. Application of the RF field leads to the formation of the triplet structure of the spin-noise spectrum, Fig.~\ref{fig:K0zz}(b). The emerging side peaks remain clearly resolved even in the presence of strong inhomogeneous broadening. Interestingly, the position and shape of the side peaks depend not only on the RF field amplitude $\Omega_{\rm rf}$ and detuning $\Omega_L - \omega_{\rm rf}$ but also on the $g$-factor dispersion, see Fig.~\ref{fig:K0zz}(c). Their spectral shift is given by $\sqrt{\Omega_{\rm rf}^2+(\bar\Omega_L-\omega_{\rm rf})^2}$ if $\delta g/g \ll |\omega_\text{rf}/\bar\Omega_L-1|$ (black dashed line) and by 
 $\Omega_{\rm rf}$ if the $g$-factor dispersion is large (thick red line).

As shown in the previous section, the correlations of the spin noise in the presence of the RF field contain
also contributions which are synchronized with the RF field and have oscillating dependence on the absolute 
time $(t+t')/2$ in addition to the dependence on the delay time $t-t'$. 
Figure~\ref{fig:K12} shows the spectra of such correlation functions $K_{xz}^{(1)}$ and $K_{xx}^{(2)}$ which
oscillate in the absolute time at the frequencies $\omega_\text{rf}$ and $2\omega_\text{rf}$, respectively.
Generally, the spectra feature the three-peak structure with the peaks of different signs and amplitudes. 
The central peak is missing in the spectrum of $K_{xz}^{(1)}$ in the resonance condition $\omega_{\rm rf} = \Omega_L$, 
Fig.~\ref{fig:K12}(a)-(c). Similarly to the spectra of $K_{xx}^{(0)}$ and $K_{zz}^{(0)}$, the dispersion of the $g$-factors leads to a modification of the side peaks and but does not affect the widths of the central peaks in the spectra of $K_{xz}^{(1)}$ and $K_{xx}^{(2)}$. Despite of the fact that the central peaks in $K_{xz}^{(1)}$ and $K_{xx}^{(2)}$ are located at $\omega = 0$, they describe the spin fluctuations oscillating in the absolute time at the frequencies $\omega_\text{rf}$ and $2\omega_\text{rf}$, respectively. Therefore, they are not obscured by a low-frequency noise and should be experimentally accessible, e.g., by means of lock-in technique.

To summarize, we have studied spin noise of an ensemble of electrons in the non-equilibrium conditions of the magnetic resonance. Interaction with the environment was taken into account as a stochastic effective magnetic field. This field leads to spin dephasing and renormalization of the Larmor precession frequency.
Application of a resonant radio-frequency magnetic field splits the peaks in the spin-noise spectra into the Mollow triplets. 
 The central peak of the triplet appears to be robust against inhomogeneous broadening, while the side peaks are smeared in such an asymmetric way that their inner wings remain sharp. Therefore, the measurements of the spin-noise spectra in the presence of radio-frequency field provide an access to the intrinsic spin relaxation time in an inhomogeneous ensemble. Apart from the modification of the  orthodox spin noise correlation functions, that depend on the time delay, the radio-frequency field also induces spin correlations that are synchronized with the radio-frequency filed and oscillate with the absolute time. They are not obscured by low-frequency noises and can be detected by means of lock-in techniques.

This work was supported by the Russian Science Foundation (project 19-12-00051). 
A.V.P. also acknowledges the support  from the Russian President Grant No. MK-599.2019.2 and the Foundation ``BASIS''.
 

%

\end{document}